\documentclass[nofootinbib,prd,a4paper,showpacs]{revtex4} 
\usepackage{amssymb,amsmath,epsfig}

\usepackage{epsfig}

\def\nn{\nonumber}        

\newcommand{\bm}[1]{\mbox{\boldmath $#1$}}

\newcommand{\open}{{<\kern -0.3 em{\scriptscriptstyle )}}}
\newcommand{\sT}{{\scriptscriptstyle T}}
\newcommand{\nslash}{\kern 0.2 em n\kern -0.45em /}
\newcommand{\Pslash}{\kern 0.2 em P\kern -0.56em \raisebox{0.3ex}{/}}
\newcommand{\pslash}{\kern 0.2 em p\kern -0.4em /}
\newcommand{\kslash}{\kern 0.2 em k\kern -0.45em /}
\newcommand{\Sslash}{\kern 0.2 em S\kern -0.56em \raisebox{0.3ex}{/}}

\newcommand{\eq}{\begin{equation}}
\newcommand{\ee}{\end{equation}}
\newcommand{\beq}{\begin{equation}}
\newcommand{\eeq}{\end{equation}}
\newcommand{\ba}{\begin{eqnarray}}
\newcommand{\ea}{\end{eqnarray}}
\newcommand{\eqa}{\begin{eqnarray}}
\newcommand{\eea}{\end{eqnarray}}

\newcommand{\sumint}{\kern 0.2 em {\textstyle\sum} \kern -1.1 em \int}

\begin{document} 

\title{TMD evolution of the Sivers asymmetry}

\author{Dani\"el Boer}
\email{d.boer@rug.nl}
\affiliation{University of Groningen,
Zernikelaan 25, NL-9747 AA Groningen, The Netherlands}

\date{\today}

\begin{abstract}
The energy scale dependence of the Sivers asymmetry in semi-inclusive deep inelastic scattering is studied numerically within the framework of 
TMD factorization that was put forward in 2011. The comparison to previous results in the literature shows that the treatment of 
next-to-leading logarithmic effects is important for the fall-off of the Sivers asymmetry with energy in the measurable regime. 
The TMD factorization based approach indicates that the peak of the Sivers asymmetry falls off with energy scale $Q$ to good approximation as 
$1/Q^{0.7}$, somewhat faster than found previously based on the first TMD factorization expressions by Collins and Soper in 1981. 
It is found that the peak of the asymmetry moves rather slowly towards higher transverse momentum values as $Q$ increases, which may be 
due to the absence of perturbative tails of the TMDs in the presented treatments. 
We conclude that the behavior of the peak of the asymmetry as a function of 
energy {\it and} transverse momentum allows for valuable tests of the TMD formalism and the considered approximations. 
To confront the TMD approach with experiment, high energy experimental data from an Electron-Ion Collider is required.
\end{abstract}


\maketitle


\section{Introduction}
The Sivers effect \cite{Sivers:1989cc,Sivers:1990fh} is a left-right asymmetry in the transverse momentum distribution of unpolarized 
quarks inside a transversely polarized proton. It is a $k_T\times S_T$ correlation in the quark transverse 
momentum ($k_T$) distribution with respect to the transverse polarization ($S_T$) of the proton moving in the $z$-direction.  
It was first defined as a transverse momentum dependent parton distribution (TMD) in \cite{Collins93}. In \cite{Boer:1997nt}
it was shown that the Sivers effect in semi-inclusive DIS (SIDIS) leads to a single transverse spin asymmetry 
$A_{UT} \sim \sin(\phi_h-\phi_S) f_{1T}^\perp D_1$, where $f_{1T}^\perp$ denotes the Sivers effect TMD
and $D_1$ the ordinary unpolarized fragmentation function. The azimuthal angles $\phi_h$ and $\phi_S$ are of the 
observed outgoing hadron's transverse momentum and of the transverse spin vector of the proton, respectively, in 
frames where the proton and virtual photon are collinear and along the $z$-direction. 
Such an asymmetry has been clearly observed in the SIDIS process by the 
HERMES \cite{Airapetian:2009ae} and COMPASS \cite{Alekseev:2010rw} experiments. Since these two experiments are 
performed at different energies ($Q$), as will future experiments at Jefferson Lab and at a possible Electron-Ion Collider (EIC), 
it is important to study the evolution of the Sivers asymmetry with energy scale $Q$. 

The evolution is dictated by TMD factorization \cite{CS81,JMY,Collins:2011zzd}. 
Explicit expressions to order $\alpha_s$ allow to obtain the leading order scale dependence of the cross section and its asymmetries. 
Evolution of the Sivers TMD and its SIDIS asymmetry has recently been studied numerically in \cite{AR} (to be referred to as AR), 
\cite{ACQR}, and \cite{APR} (APR). 
Earlier numerical studies of the evolution of the closely related Collins effect have been done in \cite{B01} (B01) and \cite{B09} (B09), 
the results of which can be carried over to the Sivers effect case upon trivial replacements. The main difference in the approaches is that 
AR and APR are based on the most recent TMD factorization expressions of \cite{Collins:2011zzd},
whereas B01 and B09 were based on the original TMD factorization expressions by Collins and Soper in 1981 \cite{CS81}. 
Although the new form of TMD factorization is preferred on theoretical grounds, because it takes care of several problematic 
issues with earlier forms (specifically, infinite rapidity divergences and divergent Wilson-line self-energies that should cancel in the 
cross section), it is not {\it a priori} clear that numerical results based on earlier expressions are invalidated or affected substantially. 
Especially for the limited energy ranges accessible in experiments previous results may still be of value and in order to judge that, 
the example of the Sivers asymmetry is considered. 

The conclusion of B01 about the energy scale dependence of the Collins single spin asymmetry (SSA) in SIDIS and by extension 
of the Sivers SSA, was that in the range $Q=30$ GeV to $Q=90$ GeV, it falls off approximately as $1/\sqrt{Q}$ (actually a range 
$1/Q^{0.5}$-$1/Q^{0.6}$ was obtained numerically, upon variation of the nonperturbative input). 
This moderate fall-off has been recently contested by APR, although no direct comparison of the same quantity was made. 
The intention here is to shed light on the differences between the 
approaches followed in B01 and B09 and the recent analyses by AR and APR, by studying one particular 
analyzing power expression. 
Although the various approaches all coincide in the double leading logarithmic approximation, in which the running of 
the strong coupling constant $\alpha_s$ is neglected, this approximation is valid only in a very small range of $Q$ values. 
The running of $\alpha_s$ will have to be included to evolve the 
Sivers asymmetry from for example the HERMES and COMPASS $Q$ scales ($\langle Q^2 \rangle = 2.4$ GeV$^2$ and $\langle Q^2 \rangle = 3.9$ GeV$^2$, respectively) to scales relevant for an EIC (with $Q$ values up to about 100 GeV). 

One conclusion of this paper will be that the approach of AR/APR yields to good approximation $1/Q^{0.7}$ for the fall-off of the peak of the asymmetry, 
when using the nonperturbative 
Sudakov factor of AR in the range from $Q \sim 3$ GeV to $Q \sim 100$ GeV, in other words is not too far from $1/Q^{0.5-0.6}$ found in B01 and even closer to what one finds using the approach in B09 (section \ref{comparison}), despite their considerably different expressions. We expect comparably moderate modifications to apply to the Collins effect asymmetries discussed in B01 and B09.  

\section{The Sivers asymmetry at tree level}

The expression for the single transverse spin asymmetry $A_{UT} \sim \sin(\phi_h-\phi_S) f_{1T}^\perp D_1$ is 
given in terms of a convolution integral:
\begin{equation} 
\frac{d\sigma(\ell H^\uparrow \to \ell' h X)}{d\Omega dx dz d^{\,2}{\bm q_\sT^{}}}=
\frac{\alpha^2\, x\, z^2\, s}{Q^4}\left(1-y+\frac{1}{2}y^2\right)\sum_{a,\bar a} e_a^2 \left\{ 
        {\cal F}\left[f_1 D_1\right] + \frac{|\bm S_{T}^{}|}{Q_T}\sin(\phi_h-\phi_{S}) 
             {\cal F}\left[\frac{\bm{q}_\sT^{}\!\cdot \!\bm{p}_\sT^{}}{M} f_{1T}^{\perp}D_1\right] + \ldots \right\}\;,
\label{crosssection}
\end{equation}
where the sum runs over all quark (and anti-quark) flavors, $e_a$ denotes the quark charge in units of the positron charge, 
and the ellipses denote contributions from other, $T$-even TMDs \cite{Boer:1999uu}.
The cross section is differential in the invariants $x=Q^2/(2\,P\cdot q)$, $z=P \cdot P_h/P\cdot q$, $y= (P \cdot q)/ (P \cdot l) \approx q^-/l^-$, 
for incoming hadron momentum $P$, outgoing hadron momentum $P_h$ and beam lepton momentum $l$ and the momentum $q$ of the virtual photon, 
defining $q^2=-Q^2$, and differential in $d\Omega = 2\,dy\,d\phi$ and $d^{\,2}{\bm q_\sT^{}}$ where $q_\sT= q +x\,P - P_h/z$, such that $Q_T^2 \equiv 
q_T^2 = - \bm q_\sT^{2} $ and $q_T$ has only transverse components in the frames where the hadrons are collinear. In the frames where the 
proton and the photon are collinear, the perpendicular component of $P_h$ satisfies: $P_{h\perp}^2 = z^2 Q_T^2$. At {\em tree level\/} 
the convolution integral with weight function $w$ is given by:
\begin{equation} 
{\cal F}\left[w\left(\bm{p}_\sT^{},\bm{k}_\sT^{}\right) f\, D\right]\equiv 
\int d^{\,2}\bm{p}_\sT^{}\; d^{\,2}\bm{k}_\sT^{}\;
\delta^2 (\bm{p}_\sT^{}+\bm q_\sT^{}-\bm 
k_\sT^{}) \, w\left(\bm{p}_\sT^{},\bm{k}_\sT^{}\right)  f^a(x,\bm{p}_\sT^2) 
D^a(z,\bm{k}_\sT^2) \;.
\end{equation}
For the evolution study we will however consider the Fourier transformed expressions, i.e.\ 
\ba
{\cal F}\left[f_1\, D_1\right]  & = &  \int \frac{d^2 \bm{b}}{(2\pi)^2} \, e^{i \bm{b} \cdot \bm{q}_T^{}} 
\, \tilde{f}_1^a(x,b^2) \, \tilde{D}_1^a(z,b^2) = \frac{1}{2\pi} \int db \, b \, J_0(bQ_T) 
\, \tilde{f}_1^a(x,b^2) \, \tilde{D}_1^a(z,b^2),\\
{\cal F}\left[\frac{\bm{q}_\sT^{}\!\cdot \!\bm{p}_\sT^{}}{M} f_{1T}^{\perp}D_1\right]  & = & -2i 
 \int \frac{d^2 \bm{b}}{(2\pi)^2} \, e^{i \bm{b} \cdot \bm{q}_T^{}} \frac{\bm{q}_\sT^{}\!\cdot \!\bm{b}^{}}{M} 
\, \tilde{f}_{1T}^{\perp\prime\; a}(x,b^2) \, \tilde{D}_1^a(z,b^2) \nn\\
& = & \frac{1}{2\pi M} \int db\, b^2 \, J_1(bQ_T) \, \tilde{f}_{1T}^{\perp \prime \; a}(x,b^2) \, \tilde{D}_1^a(z,b^2),
\ea
where $\tilde{f}_1(x,\bm{b})$ and $\tilde{D}_1(z,\bm{b})$ denote the Fourier transforms of the unpolarized TMD distribution function 
$f_1(x,\bm{p}_\sT^2)$ and TMD fragmentation function ${D}_1(z,{\bm{k}_\sT^2})$. We have also defined 
\beq
-2i b^\alpha \tilde{f}_{1T}^{\perp \prime \; a}(x,b^2) \equiv \int \frac{d^2 \bm{p}_{\sT}^{}}{(2\pi)^2} e^{i\bm{p}_{\sT}^{} \cdot \bm{b}} p_\sT^\alpha {f}_{1T}^{\perp \; a}(x,p_\sT^2).
\eeq
This yields for the analyzing power of the $\sin(\phi_h-\phi_S)$ asymmetry 
\ba
A_{UT}(Q_T) & = & \frac{x\, z^2\,\left(1-y+\frac{1}{2}y^2\right) \sum_{a,\bar a} e_a^2  
             {\cal F}\left[\bm{q}_\sT^{}\!\cdot \!\bm{p}_\sT^{} f_{1T}^{\perp}D_1\right]}{x\, z^2\, \left(1-y+\frac{1}{2}y^2\right) M Q_T \sum_{b,\bar b} e_b^2 {\cal F}\left[f_1 D_1\right]} \nn\\
& = & \frac{x\, z^2\,\left(1-y+\frac{1}{2}y^2\right) \sum_{a,\bar a} e_a^2 \int db\,b^2 \,J_1(bQ_T) 
\,\tilde{f}_{1T}^{\perp\prime \; a}(x,b^2) \, \tilde{D}_1^a(z,b^2)}{x\, z^2\, \left(1-y+\frac{1}{2}y^2\right)M Q_T  \sum_{b,\bar b} e_b^2 \int db\, b \, J_0(bQ_T) 
\, \tilde{f}_1^b(x,b^2) \, \tilde{D}_1^b(z,b^2)}  .          
\ea
Keeping in mind that the flavor indices in numerator and denominator are part of separate summations, we define 
\beq
{\cal A}_{ab}(x,z,Q_T) \equiv 
\frac{\int db\, b^2 \,J_1(bQ_T) 
\,\tilde{f}_{1T}^{\perp\prime \; a}(x,b^2) \, \tilde{D}_1^a(z,b^2)}{M Q_T \int db\, b \, J_0(bQ_T) 
\, \tilde{f}_1^b(x,b^2) \, \tilde{D}_1^b(z,b^2)} 
\label{calAQTab}
\eeq
At tree level this would be the relevant asymmetry quantity at all energy scales. Beyond tree level it would be a valid 
expression at one particular scale only. 
The expressions at other scales can then be obtained by evolution of the parameters involved.  
To discuss this in more detail, we will first discuss the relevant aspects of TMD factorization.  

\section{\label{sec:Scale}Scale dependence of the TMD factorized cross section} 

The proof of TMD factorization of processes such as semi-inclusive DIS or Drell-Yan, has recently been 
finalized \cite{Collins:2011zzd,Collins:2011ca}. It involves a new definition of TMDs that incorporates
the soft factor, which then no longer appears explicitly in the cross section expression.
In this TMD formalism the differential cross section of for instance the SIDIS process at small
$Q_T^2/Q^2$ is written as
\beq
\frac{d\sigma}{dx dy dz d\phi {d^2 \bm{q}_{\sT}^{}}} = \int d^2 {b} 
\, e^{-i {\bm{b} \cdot \bm{q}_T^{}}} \tilde{W}({\bm{b}}, Q;
  x, y, z) + {\cal O}\left(Q_T^2/Q^2\right). 
\label{CS81xs}
\eeq
The integrand for unpolarized hadrons and unpolarized quarks of flavor $a$ is given by
\ba
\tilde{W}({\bm{b}}, Q; x,y,z) & = &
{\sum_{a}\,\tilde{f}_1^{a}(x,{\bm{b}};\zeta_F, \mu)}
{\tilde{D}_1^{a}(z,{\bm{b}};\zeta_D,\mu)} H\left(y,Q;\mu\right),
\label{CS81W}
\ea
 The partonic hard scattering part $H$, 
which for the choice $\mu = Q$ takes the form
\ba 
H\left(Q;\alpha_s(Q)\right) & \propto & e_a^2 \left(1 + \alpha_s(Q^2) F_1 + {\cal O}(\alpha_s^2) \right),
\label{HatQ}
\ea
where $F_1$ denotes a renormalization-scheme-dependent finite term. 
The dependence of the TMDs on $\zeta_F, \zeta_D$ and $\mu$ will be discussed in detail next. 
Once the factorization expression is given, with all its scale dependence, the evolution of 
TMD cross sections can be obtained. To obtain the evolution of spin asymmetries in these cross sections, 
one has to include spin dependent TMDs, whose Fourier transforms can be odd under $\bm{b} \to - \bm{b}$, 
such as the Sivers or Collins effect.  

Note that Eq.\ (\ref{CS81xs}) contains no integrals over the {\it partonic} momentum fractions, 
which only appear in the large $Q_T$ (or equivalently, small $b$) limit. What is considered large $Q_T$ depends on $Q$. 
For asymptotically large $Q$, the main contribution to the $b$ integral is from small $b$ values, which means that 
the $b$ dependence of the (Fourier transformed) TMDs can be calculated entirely perturbatively. For the $Q$ values of 
HERMES and COMPASS, this is certainly not the case. The {\em peak} of the Sivers asymmetry for $Q$ 
values in the range $3$ to $100$ GeV will be located at $Q_T$ values for which the $b$-integration receives important contributions 
from $b$ values that do not allow for a perturbative calculation of the $b$ dependence. This will require separate treatment 
of the small and large $b$ regions.    

\subsection{Sudakov factor} 

In Eq.\ (\ref{CS81W}) the Fourier transformed TMDs $\tilde f_1$ and the hard part $H$ have a dependence on the 
renormalization scale $\mu$. It will be chosen $\mu=Q$, such that 
there are no $\ln Q/\mu$ terms in the hard part, as in Eq.\ (\ref{HatQ}). 
In order to avoid large logarithms, the TMDs will be taken at either the scale $\mu_b=C_1/b=2e^{-\gamma_E}/b$ ($C_1 \approx 1.123$), 
or at the fixed scale $Q_0$ which is to be taken as the lowest scale for which perturbation theory is expected to be
trustworthy (a common choice is $Q_0=1.6$ GeV). Evolving the TMDs from the scale $Q$ to $\mu_b$ or $Q_0$ 
will result in a separate factor, called the Sudakov factor, which will be discussed in this subsection. We will primarily focus on the fixed 
scale case, such that the TMDs are always considered at the same scale $Q_0$ when integrating over $b$. This has the advantage, 
as will become clear below, that the remaining $b$-dependence of the TMDs is perturbatively calculable.
The fixed scale option was already suggested by Collins \& Soper \cite{CS81} and explicitly used by Ji {\em et al.} \cite{JMY,Idilbi}, 
who called it $\mu_L$, and in B09 and APR. It is not necessarily always the optimal choice though, which depends on the absence of 
large logarithmic corrections. For the specific quantity and the energy and momentum region considered here, it appears to be an 
appropriate choice, that moreover will allow us to compare the approaches of APR and B09 more directly.  

The TMDs also depend on $\zeta_{F(D)}$, which are defined as \cite{ACQR}:
\beq
\zeta_F = M^2 x^2 e^{2(y_{P}-y_s)},\quad 
\zeta_D = M_{h}^2 e^{2(y_s-y_{h})}/z^2,
\eeq
where $y_{P(h)}$ denotes the rapidity of the incoming proton and outgoing hadron and 
the dependence on the arbitrary rapidity cut-off $y_s$ cancels in the cross section, 
where only the product $\zeta_F \zeta_D \approx Q^4$ enters. 

The evolution of the TMD $\tilde f$ in both $\zeta$ and $\mu$ is known and given by the following Collins-Soper and Renormalization Group equations, respectively \cite{Collins:2011zzd}:
\ba
\frac{d \ln \tilde{f}(x,b;\zeta,\mu)}{d \ln \sqrt \zeta} & = & \tilde K(b;\mu),\\
\frac{d \ln \tilde{f}(x,b;\zeta,\mu)}{d\ln \mu} & = & \gamma_F(g(\mu);\zeta/\mu^2),
\ea
where $d\tilde K/d\ln \mu = -\gamma_K(g(\mu))$ and $\gamma_F(g(\mu);\zeta/\mu^2)= 
\gamma_F(g(\mu);1)-\frac{1}{2}\gamma_K(g(\mu))\ln(\zeta/\mu^2)$. With these evolution equations 
one can evolve the TMDs to the scale $\mu_b$ or $Q_0$, i.e.\
\ba
\tilde{f}(x,b; \zeta, Q) & = & \tilde{f}(x,b; \mu_b^2, \mu_b) \exp \bigg\{\ln\left(\frac{\sqrt{\zeta}}{\mu_b}\right)\tilde K(b,\mu_b) +
\int_{\mu_b}^Q \frac{d\mu}{\mu} \big[ \gamma_F(g(\mu);1)-\ln\left(\frac{\sqrt{\zeta}}{\mu}\right) \gamma_K(g(\mu)) \big] \bigg\},
\label{fmub}
\ea
or
\ba
\tilde{f}(x,b^2; \zeta, Q) & = & \tilde{f}(x,b^2; Q_0^2,Q_0) \exp \bigg\{\ln\left(\frac{\sqrt{\zeta}}{Q_0}\right)\tilde K(b,Q_0) +
\int_{Q_0}^Q \frac{d\mu}{\mu} \big[ \gamma_F(g(\mu);1)-\ln\left(\frac{\sqrt{\zeta}}{\mu}\right) \gamma_K(g(\mu)) \big] \bigg\}.
\label{fQ0}
\ea
The latter expression is however not optimal\footnote{The author 
is grateful to John Collins and Ted Rogers for pointing this out.}, since it does not take care of possible large logarithms in $\mu_b/Q_0$. 
It is more appropriate to use instead \cite{APR}:
\ba
\tilde{f}(x,b^2; \zeta, Q) & = & \tilde{f}(x,b^2; Q_0^2,Q_0) \exp \bigg\{\ln\left(\frac{\sqrt{\zeta}}{Q_0}\right)\left(\tilde K(b,\mu_b) +
\int_{Q_0}^{\mu_b} \frac{d\mu}{\mu} \gamma_K(g(\mu)) \right) \nn\\
&& + \int_{Q_0}^Q \frac{d\mu}{\mu} \big[ \gamma_F(g(\mu);1)-\ln\left(\frac{\sqrt{\zeta}}{\mu}\right) \gamma_K(g(\mu)) \big] \bigg\}.
\label{fQ0used}
\ea
Similar equations can be obtained for the TMD fragmentation functions $D$, cf.\ \cite{AR}.
This yields the expressions:
\beq
\tilde{f}_1^a(x,b^2; \zeta_F, \mu) \, 
\tilde{D}_1^b(z,b^2;\zeta_D, \mu)=  e^{-S(b,Q)} \tilde{f}_1^a(x,b^2; \mu_b^2, \mu_b) \, 
\tilde{D}_1^b(z,b^2;\mu_b^2, \mu_b),
\eeq
with 
\ba
S(b,Q) & = & - \ln\left(\frac{Q^2}{\mu_b^2}\right)\tilde K(b,\mu_b) - \int_{\mu_b^2}^{Q^2} \frac{d\mu^2}{\mu^2} \big[ \gamma_F(g(\mu);1)-\frac{1}{2} \ln\left(\frac{Q^2}{\mu^2}\right) \gamma_K(g(\mu))\big],
\label{Smubform}
\ea
and 
\beq 
\tilde{f}_1^a(x,b^2; \zeta_F, \mu) \, 
\tilde{D}_1^b(z,b^2;\zeta_D, \mu) = e^{-S(b,Q,Q_0)} \tilde{f}_1^a(x,b^2; Q_0^2, Q_0) \, 
\tilde{D}_1^b(z,b^2;Q_0^2, Q_0),
\eeq
with 
\beq
S(b,Q,Q_0)  =  - \ln\left(\frac{Q^2}{Q_0^2}\right)\left(\tilde K(b,\mu_b) +
\int_{Q_0}^{\mu_b} \frac{d\mu}{\mu} \gamma_K(g(\mu)) \right) - \int_{Q_0^2}^{Q^2} \frac{d\mu^2}{\mu^2} \big[ \gamma_F(g(\mu);1)-\frac{1}{2} \ln\left(\frac{Q^2}{\mu^2}\right) \gamma_K(g(\mu))\big],
\label{SQ0form}
\eeq
where we have used that $\gamma_D=\gamma_F$ to the order in $\alpha_s$ considered here. 

\subsection{Perturbative Sudakov factor} 
The various quantities in the Sudakov factor to order $\alpha_s$ are given by \cite{CS81,AR}:
\ba
\tilde{K}(b,\mu) & = & -\alpha_s(\mu)\frac{C_F}{\pi} \ln(\mu^2 b^2/C_1^2) + {\cal O}(\alpha_s^2), \label{Kmu}\\
\gamma_K(g(\mu)) & = & 2\alpha_s(\mu)\frac{C_F}{\pi}+ {\cal O}(\alpha_s^2),\\
\gamma_F(g(\mu), \zeta/\mu^2) & = & \alpha_s(\mu)\frac{C_F}{\pi}\left(\frac{3}{2}- \ln\left(\zeta/\mu^2\right) \right) + {\cal O}(\alpha_s^2). 
\label{gammaF}
\ea
Here it should be noted that for a running $\alpha_s$, the choice of scale $\mu$, and hence of the integration range over $\mu$ in the Sudakov factor, matters much for the size of the errors here generically denoted by ${\cal O}(\alpha_s^2)$. Depending on the choice of factorized expression, including the choice of using $\mu_b$ or $Q_0$, the error in the final result for the asymmetry may vary considerably in size. We emphasize that the fixed scale $Q_0$ choice is not necessarily the optimal choice in all cases.  
 
From these perturbative expressions one obtains the following perturbative Sudakov factors:
\ba
S_p(b,Q) & = & \frac{C_F}{\pi} \int_{\mu_b^2}^{Q^2} \frac{d\mu^2}{\mu^2} \alpha_s(\mu)\left(\ln \frac{Q^2}{\mu^2}- \frac{3}{2} \right) + {\cal O}(\alpha_s^2), \label{SpbQ}\\
S_p(b,Q,Q_0) & = & - \frac{C_F}{\pi} \ln\left(\frac{Q^2}{Q_0^2}\right) \int_{Q_0^2}^{\mu_b^2} \frac{d\mu^2}{\mu^2} \alpha_s(\mu) 
+  \frac{C_F}{\pi} 
\int_{Q_0^2}^{Q^2}  \frac{d\mu^2}{\mu^2} \alpha_s(\mu)\left(\ln\frac{Q^2}{\mu^2}-\frac{3}{2} \right)+ {\cal O}(\alpha_s^2).
\label{SpbQQ0}
\ea
Including the one-loop running of $\alpha_s$ one can perform the $\mu$ integrals explicitly. We define
\beq
S_1(Q,Q_0) \equiv \frac{C_F}{\pi}  \int_{Q_0^2}^{Q^2}  \frac{d\mu^2}{\mu^2} \alpha_s(\mu) \ln \frac{Q^2}{\mu^2}
= -\frac{16}{33-2n_f} \left[ \ln\left(\frac{Q^2}{Q_0^2}\right)+
\ln\left(\frac{Q^2}{\Lambda^2}\right)\; \ln\left[1- \frac{\ln\left( 
Q^2/Q_0^2\right)}{\ln\left(Q^2/\Lambda^2\right)} \right]\right], 
\eeq
such that (dropping non-logarithmic finite terms)
\ba
S_p(b,Q,Q_0) & = & S_1(Q,Q_0) -  \frac{16}{33-2n_f} \ln\left(\frac{Q^2}{Q_0^2}\right) \ln \left[\frac{\ln\left(  
\mu_b^2/\Lambda^2\right)}{\ln\left(Q_0^2/\Lambda^2\right)} \right].
\label{SpbQQ0used}
\ea

The above expressions for the Sudakov factor are valid in the perturbative region $b<1/Q_0$. Strictly speaking, at {\em very\/} small $b$ $(<
1/Q)$ the perturbative expressions do not have the correct behavior $S\to 0$ in the limit $b \to 0$. If this region gives 
important contributions, it requires modifications of the expressions (for example the regularization discussed in \cite{ParisiPetronzio}) 
or else it can lead to artifacts, such as `Sudakov enhancement' of the cross section. 
For the Sivers asymmetry calculation we find that such a modification 
is needed if one uses Eq.\ (\ref{fQ0}) instead of Eq.\ (\ref{fQ0used}). Using Eq.\ (\ref{fQ0used}), and hence Eq.\ (\ref{SpbQQ0used}), leads to only 
minor contributions from the region $b<1/Q$, so we will use them without modifications. 

\subsection{Nonperturbative Sudakov factor} 
As said, the above expressions for the Sudakov factor are valid in the perturbative region $b<1/Q_0$. Since at sub-asymptotic $Q$ values, 
the Fourier transform involves also the nonperturbative 
region of large $b$ and we explicitly focus on the region $Q_T^2 \ll Q^2$, we will have to deal with $b>1/Q_0$ also. This can be
done for instance via the introduction of a $b$-regulator 
\cite{CSS-85}\footnote{An alternative method using a deformed contour in the complex $b$ plane has been put forward in 
Ref.\ \cite{Laenen:2000de}.}: 
$b \to b_*=b/\sqrt{1+b^2/b_{\max}^2}$, such that $b_*$ is always smaller
than $b_{\max}$. 
One then rewrites $\tilde{W}(b)$ as \cite{CSS-85}:
\beq
\tilde{W}(b) \equiv \tilde{W}(b_*) \, e^{-S_{NP}(b)} ,
\eeq
where for the function 
$\tilde{W}(b_*)$ a perturbative expression can be used. 
The nonperturbative Sudakov factor $S_{NP}$ to be used here is the one from Aybat and Rogers \cite{AR}
\beq
S_{NP}(b,Q,Q_0) = \left[g_2 
\ln\frac{Q}{2Q_0} +g_1\left(1+2g_3 \ln \frac{10xx_0}{x_0+x}\right)\right] b^2 ,
\label{ARSNP}
\eeq
with $g_1=0.201\,\text{GeV}^{2}, g_2=0.184\,\text{GeV}^{2}, g_3=-0.13, x_0=0.009,  
Q_0=1.6 \,\text{GeV}$ and $b_{\max}= 1.5 \,\text{GeV}^{-1}$. This form is chosen such that at low energy ($Q=\sqrt{2.4}$ GeV) it yields a Gaussian with $\langle k_T^2 \rangle=0.38$ GeV$^2$ that resulted from fits to SIDIS data \cite{Schweitzer:2010tt} and at high energy it matches onto the form fitted simultaneously to Drell-Yan and $Z$ boson production data \cite{Landry:2002ix}. Like in AR, $x=0.09$ will be chosen, resulting in 
\beq
S_{NP}(b,Q,Q_0) = \left[0.184 \ln\frac{Q}{2Q_0}+0.332 \right] b^2 .
\label{actualSNP}
\eeq
Note that $S_{NP}$ has a $Q^2$ dependence that is in accordance with the phenomenological observation that the average partonic transverse momentum grows as the energy ($Q$ or $\sqrt{s}$) increases (cf.\ Fig.\ 12.3 of \cite{Begel:1999rc}). 
The $Q$-independent part of $S_{NP}$ can in general be spin dependent, which means that strictly speaking one should allow for a somewhat 
different $S_{NP}$ in the numerator and denominator of the Sivers asymmetry. At lower $Q^2$ this can become relevant. 
Although quantitatively the results depend considerably on the choice of 
$S_{NP}$, in B01/B09 it was found that the $Q^2$ dependence of the ratio is not very sensitive to it. Below we will briefly comment on it further.  

\subsection{TMDs at small $b$}
Using the above perturbative and nonperturbative Sudakov factors, we end up with the asymmetry expression: 
\beq
{\cal A}_{ab}(x,z,Q_T) \equiv 
\frac{\int db\, b^2\, J_1(bQ_T) 
\,\tilde{f}_{1T}^{\perp\prime \; a}(x,b_*^2;Q_0^2,Q_0)\, \tilde{D}_1^a(z,b_*^2;Q_0^2,Q_0) \exp\left({-S_p(b_*,Q,Q_0)-S_{NP}(b,Q/Q_0)}\right) 
}{M Q_T \int db\, b\, J_0(bQ_T) \tilde{f}_1^b(x,b_*^2;Q_0^2,Q_0) \tilde{D}_1^b(z,b_*^2;Q_0^2,Q_0)\exp\left({-S_p(b_*,Q,Q_0)-S_{NP}(b,Q/Q_0)}\right)} . 
\label{calAQTabs}
\eeq
Assuming the TMDs are slowly varying as a function of $b_*$, as they are when the TMDs are taken to be (broad) Gaussians, like in B01/B09\footnote{Note that in B01/B09 the Gaussian width of the Sivers TMD appears in the asymmetry expressions, because of the derivative in ${f}_{1T}^{\perp\prime \; a}(x;Q_0)$.}, yields
\ba
{\cal A}_{ab}(x,z,Q_T) & = & \frac{f_{1T}^{\perp\prime \; a}(x;Q_0) D_1^a(z;Q_0)}{M^2 f_{1}^{b}(x;Q_0) D_1^b(z;Q_0)} {\cal A}(Q_T) 
\label{calAQTapprox}
\ea
for some functions $f_{1}(x;Q_0)$, $D_1(z;Q_0)$ ({\it a priori} not coinciding with the collinear parton distribution and fragmentation functions at the scale $Q_0$), $f_{1T}^{\perp\prime \; a}(x;Q_0)$, and  
\ba
{\cal A}(Q_T) & \equiv & M \frac{\int db\, b^2 \, J_1(bQ_T) \,\exp\left({-S_p(b_*,Q,Q_0)-S_{NP}(b,Q/Q_0)}\right)}{\int db\, b \, J_0(bQ_T) 
\, \exp\left({-S_p(b_*,Q,Q_0)-S_{NP}(b,Q/Q_0)}\right)} .
\label{calAQTs}
\ea
The approach of AR includes the perturbative expansion of the $b_*$ dependence of the TMDs, which yields 
$\ln Q_0^2 b_*^2$ terms, but also integrals over momentum fractions and mixing between quark and gluon operators. Clearly this is the more sophisticated approach, but it also makes it harder to handle. It was not included in the analysis of APR which confronts the Sivers asymmetry evolution 
with HERMES and COMPASS data. Here we will also not include it. The expression for ${\cal A}_{ab}(x,z,Q_T)$ in Eq.\ (\ref{calAQTapprox}) thus corresponds to the recent TMD factorization based approach discussed in APR. The approach followed in B09 included some $b_*$ dependence beyond that of the Sudakov factor, arising from the so-called soft factor, see Sec.\ \ref{comparison}. It was found that it makes the asymmetry fall off somewhat faster with energy. 

It should be noted that the above simplification of dropping the perturbative tails of the TMDs (the $b_*$ dependence of the TMDs), every asymmetry involving one $k_T$-odd TMD will be of the form in Eq.\ (\ref{calAQTapprox}), and hence proportional to ${\cal A}(Q_T)$. This has the advantage that the results obtained below also apply to for instance the Collins asymmetry in SIDIS. Of course, the latter will be multiplied by a different $x$ and $z$ dependent prefactor. 

\section{Numerical study of the $Q_T$ and $Q$ dependence of ${\cal A}(Q_T)$} 
\begin{figure}[htb]
\begin{center}
\includegraphics[height=5 cm]{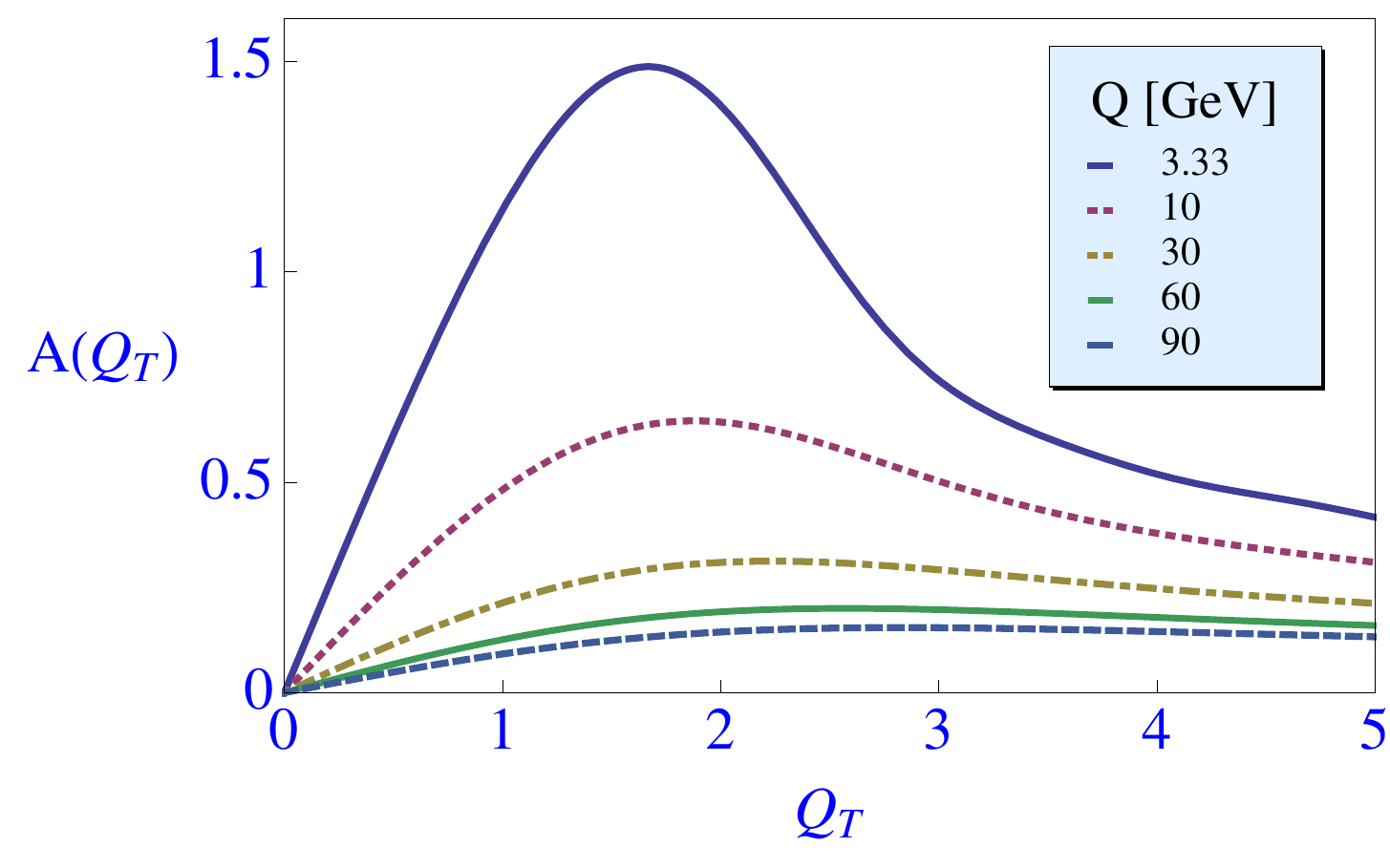}\hspace{1 cm}
\includegraphics[height=5 cm]{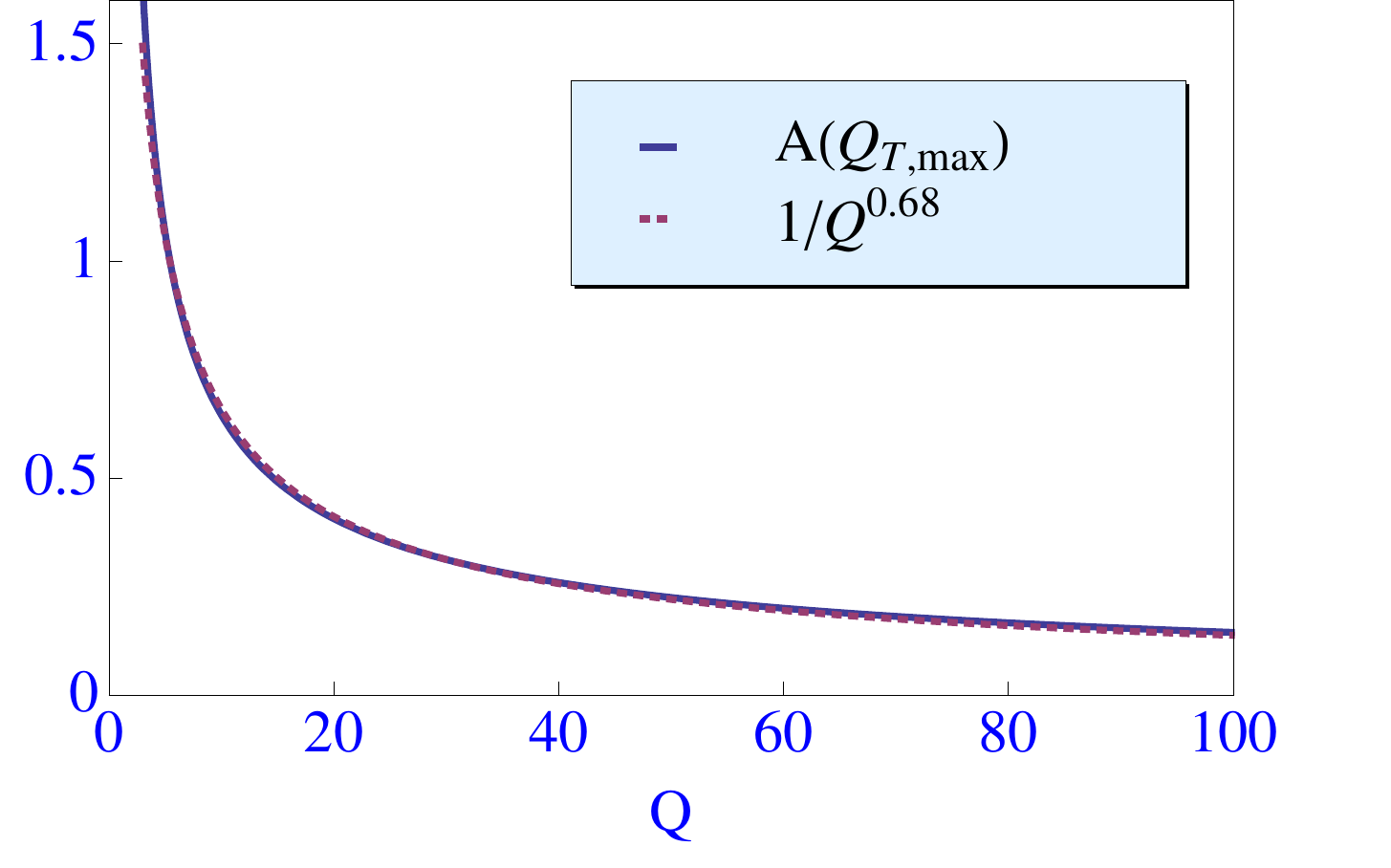}
\caption{Left: the TMD factorization based asymmetry factor ${\cal A}(Q_T)$ (in units of
$M^2$) at $Q=3.33, 10, 30, 60, 90 \, \text{GeV}$. Right: the asymmetry factor ${\cal A}(Q_T)$ evaluated at $Q_{T,\max}$ (solid line) plotted as function of $Q$  and compared to a line (dotted) with $1/Q^{0.68}$ fall-off, constructed to coincide at the point $Q=30 \, \text{GeV}$.}
\label{SplotQT}
\end{center}
\vspace{-2 mm}
\end{figure}

The expression studied numerically is ${\cal A}(Q_T)$ as given in Eq.\ (\ref{calAQTs}), using the perturbative Sudakov factor in Eq.\ (\ref{SpbQQ0used}) and the nonperturbative Sudakov factor $S_{NP}$ from AR/APR in Eq.\ (\ref{actualSNP}). In Fig.\ \ref{SplotQT} (left) ${\cal A}(Q_T)$ is plotted for various energies. As can be seen, the asymmetry has a single peak structure, whose magnitude falls off with energy. 
In Fig.\ \ref{SplotQT} (right)  ${\cal A}(Q_T)$ evaluated at $Q_T=Q_{T,\max}$ at which the asymmetry reaches its peak, is plotted as a function of $Q$ and compared to a very simple power law approximation. This shows that the peak of ${\cal A}(Q_T)$ has to good approximation a $1/Q^{0.68}$ fall-off, 
which is only slightly faster than the results of B01, where a $1/Q^{0.5-0.6}$ dependence was found. 

\begin{figure}[htb]
\begin{center}
\includegraphics[height=5 cm]{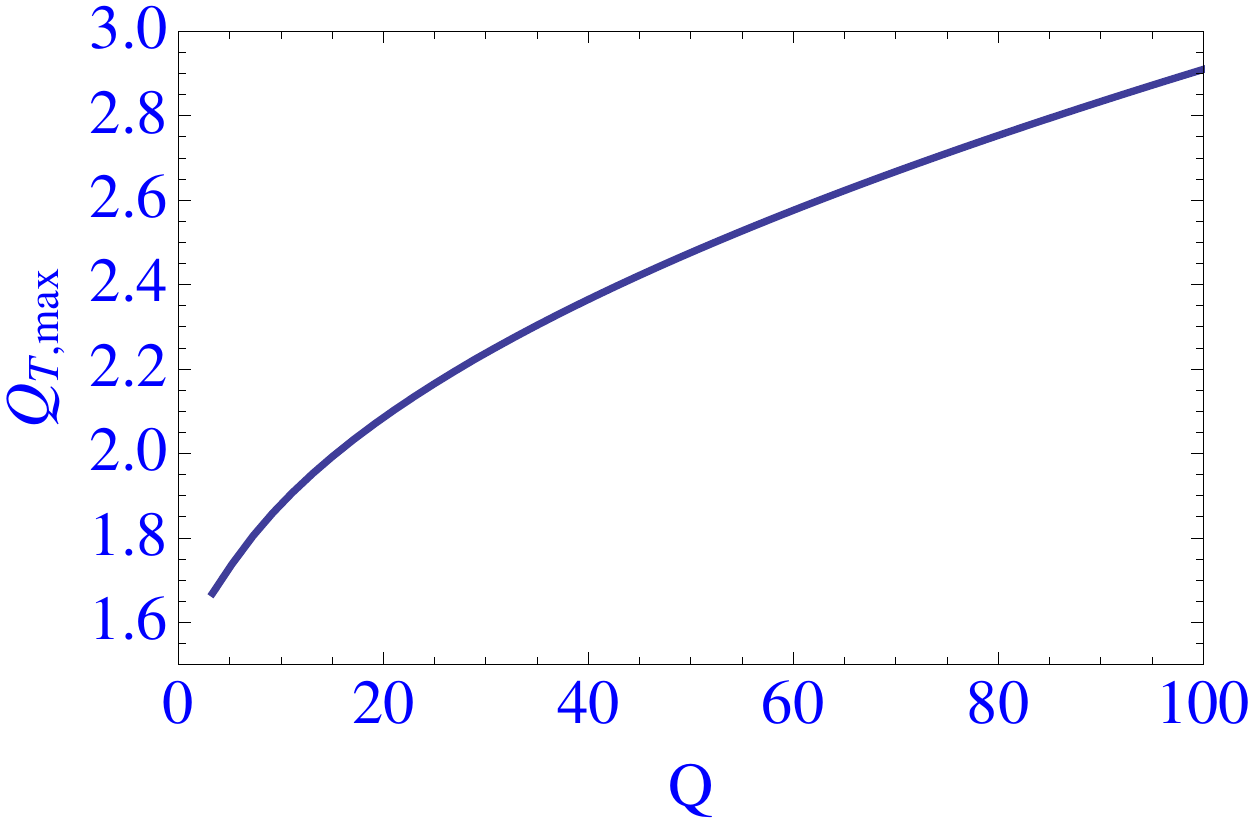}
\caption{The peak position $Q_{T,\max}$ as function $Q$.}
\label{QTmaxvsQ}
\end{center}
\vspace{-2 mm}
\end{figure}

The reason for focusing on the $Q$ dependence of the peak rather than of the asymmetry at a fixed transverse momentum or of an integral of the asymmetry, 
is simply that for future experiments at higher scales one is first of all interested to know the minimal sensitivity required to observe a nonzero asymmetry signal.  
It is thus most interesting to study the evolution in the $Q_T$ region where the asymmetry is largest, which is a region that shifts towards higher transverse momentum values as the energy increases. It matters less for experimental studies that on the sides of the peak where the asymmetry is considerably smaller, it falls off even faster. In Fig.\ \ref{QTmaxvsQ} it is shown how $Q_{T,\max}$ moves towards higher values as $Q$ increases. 
If this growth of $Q_{T,\max}$ is not confirmed experimentally, it likely means that the $b_*$ treatment of the TMDs is oversimplified, i.e.\ that the perturbative tails of the TMDs matter. 
Also the choice of $S_{NP}$ may make a difference. To give an idea of the dependence of the results on $S_{NP}$: multiplying $S_{NP}$ in Eq.\ (\ref{actualSNP}) by a factor of 2 yields $1/Q^{0.64}$ and by a factor of $1/2$ yields $1/Q^{0.75}$. Although the power of the fall-off is not affected much, the peak and its position do change considerably, in general by 40-50\% (very similar to what was found in B01). It should be kept in mind though that the $S_{NP}$ used here is fitted to available unpolarized data over the entire considered $Q^2$ range and is therefore very appropriate for the denominator of the asymmetry. Variations by a factor of 2 are thus not realistic (remember that they appear in an exponent). But what is very well possible is that the $Q^2$ independent part of $S_{NP}$ in the numerator is smaller than what is used here (larger is not allowed by the positivity bound at $Q=2Q_0$). If it is reduced by a factor $1/2$ it yields $1/Q^{0.79}$. Taking into account the uncertainty from $S_{NP}$, we thus estimate the power to be in the range 0.6-0.8.  

\section{\label{comparison}Comparison to a CS factorization based approach}
We already compared the results obtained within the recent TMD factorization approach with some results of the study B01. 
Since the approach of B09 was an improved version of B01, also based on the original Collins-Soper (CS) factorization, it may be  
of interest to compare the above results to those that would follow from B09. This can be useful for estimating the size of the expected 
modifications of old results for other asymmetries as well.  

The approach in B09 was based on a fixed scale $Q_0$ version of the CS factorization, obtained from Ref.\ \cite{CS81} by 
using the relevant renormalization group equations. This resulted in a perturbative Sudakov factor \cite{B09}: 
\ba 
S_{B09}(b,Q,Q_0) & = & C_F \int_{Q_0^2}^{Q^2} \frac{d \mu^2}{\mu^2}
\frac{\alpha_s(\mu)}{\pi} \, \left[ \ln \frac{Q^2}{{\mu}^2} + 
\ln Q_0^2 b^2 + F_3 \right],
\ea
where the renormalization-scheme-dependent finite term $F_3=2\gamma_E-3/2-\ln 4$ will be dropped. We emphasize that although this expression is here denoted by $S_{B09}$, because it was written in this way in B09, it can straightforwardly be obtained from the original CS paper by replacing its 
$\mu_1=C_1/b \to Q_0$ and carrying through all the relevant subsequent replacements in the various renormalization group equations. 

This Sudakov factor can be rewritten as 
\ba
S_{B09}(b,Q,Q_0) & = & S_1(Q,Q_0) + \frac{C_F}{\pi} \ln Q_0^2 b^2  \int_{Q_0^2}^{Q^2} \frac{d \mu^2}{\mu^2} \alpha_s(\mu),
\label{SB09used}
\ea
which differs from $S_p(b,Q,Q_0)$ in Eq.\ (\ref{SpbQQ0used}) only by sub-leading logarithmic terms. In the leading (double) logarithmic 
approximation of fixed coupling constant, all perturbative Sudakov expressions coincide with the well-known result of \cite{ParisiPetronzio}. 
Any numerical difference between the approaches thus indicates sensitivity to single logarithms, like from the running of $\alpha_s$. 
Needless to say, this becomes more important the larger the $Q$ range is in the comparison.  

In the expressions of B09 also a soft term $\tilde{U}(b; Q_0,\alpha_s(Q_0))$ 
needs to be included \cite{B09}: 
\ba
\tilde U(b;Q_0,\alpha_s(Q_0)) & = & 1 - \frac{\alpha_s(Q_0^2)}{\pi}
C_F \left( \ln Q_0^2 b^2 + F_2 \right) + {\cal O}(\alpha_s^2),
\label{leadingtermu}
\ea
where the finite term $F_2=\ln \pi + \gamma_E$ will be dropped again. All this amounts to inserting in Eq.\ (\ref{calAQTabs}):
\ba 
\tilde{f}_{1T}^{\perp\prime \; a}(x,b_*^2;Q_0^2,Q_0)\tilde{D}_1^a(z,b_*^2;Q_0^2,Q_0) & = & {f}_{1T}^{\perp\prime \; a}(x;Q_0) {D}_1^a(z;Q_0) \tilde{U}(b_*; Q_0,\alpha_s(Q_0)),\\ 
\tilde{f}_1^b(x,b_*^2;Q_0^2,Q_0) \tilde{D}_1^b(z,b_*^2;Q_0^2,Q_0) & = & {f}_{1}^{b}(x;Q_0) {D}_1^b(z;Q_0) \tilde{U}(b_*; Q_0,\alpha_s(Q_0)),
\ea
and replacing $S_p$ by $S_{B09}$.
The approach followed in B09 thus includes some $b_*$ dependence beyond that of the Sudakov factor. 
As said, this leads to a somewhat faster fall-off of the asymmetry with energy. 

Putting this together, the expression to be evaluated is: 
\ba
{\cal A}_{B09}(Q_T) & \equiv & M \frac{\int db\, b^2 \, J_1(bQ_T) \,\tilde{U}(b_*; Q_0,\alpha_s(Q_0))\, \exp\left({-S_{B09}(b_*,Q,Q_0)-S_{NP}(b,Q/Q_0)}\right)}{\int db\, b \, J_0(bQ_T) 
\, \tilde{U}(b_*; Q_0,\alpha_s(Q_0))\, \exp\left({-S_{B09}(b_*,Q,Q_0)-S_{NP}(b,Q/Q_0)}\right)} .
\label{calAQTB09}
\ea

\begin{figure}[htb]
\begin{center}
\includegraphics[height=5 cm]{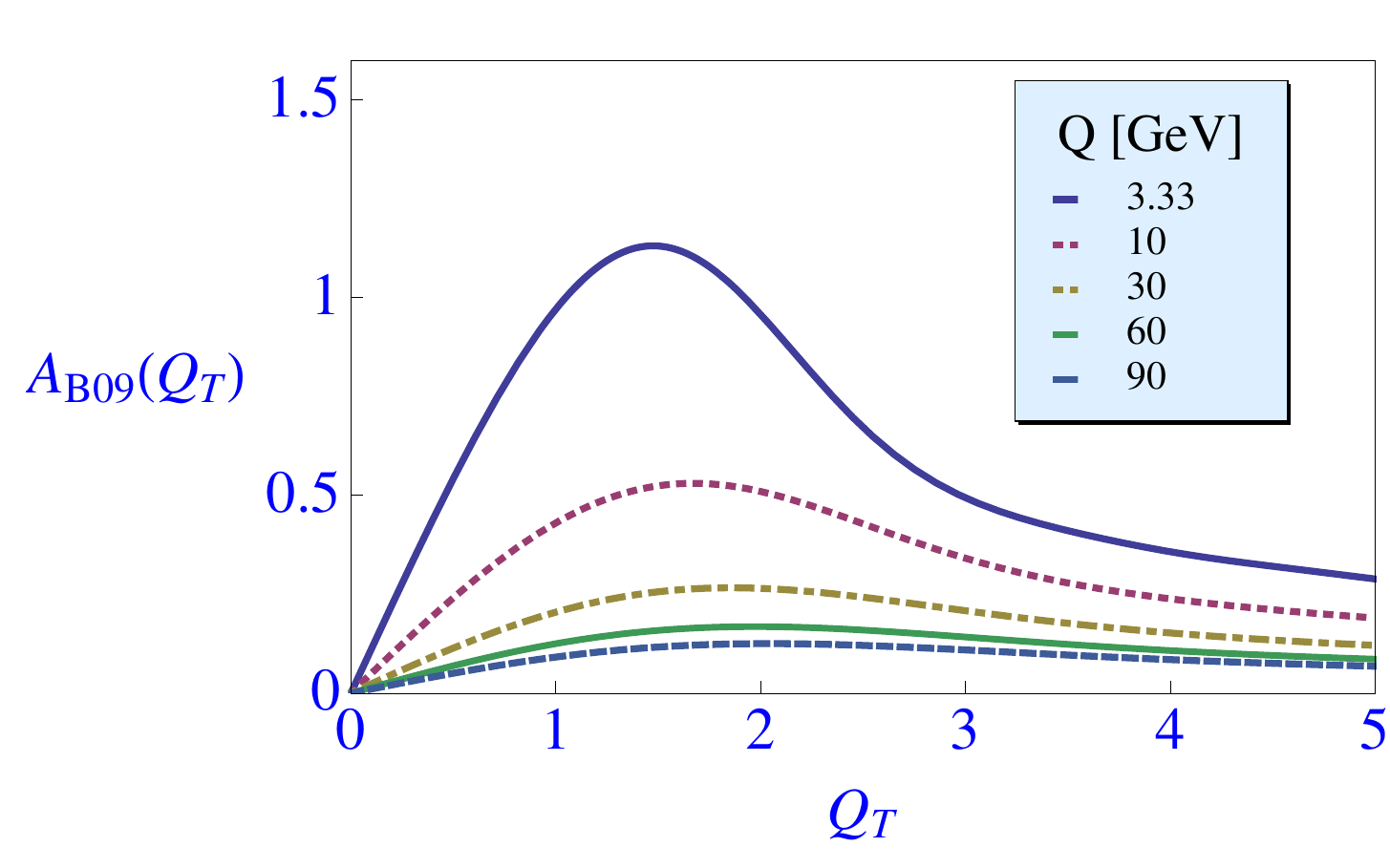}\hspace{1 cm}
\includegraphics[height=5 cm]{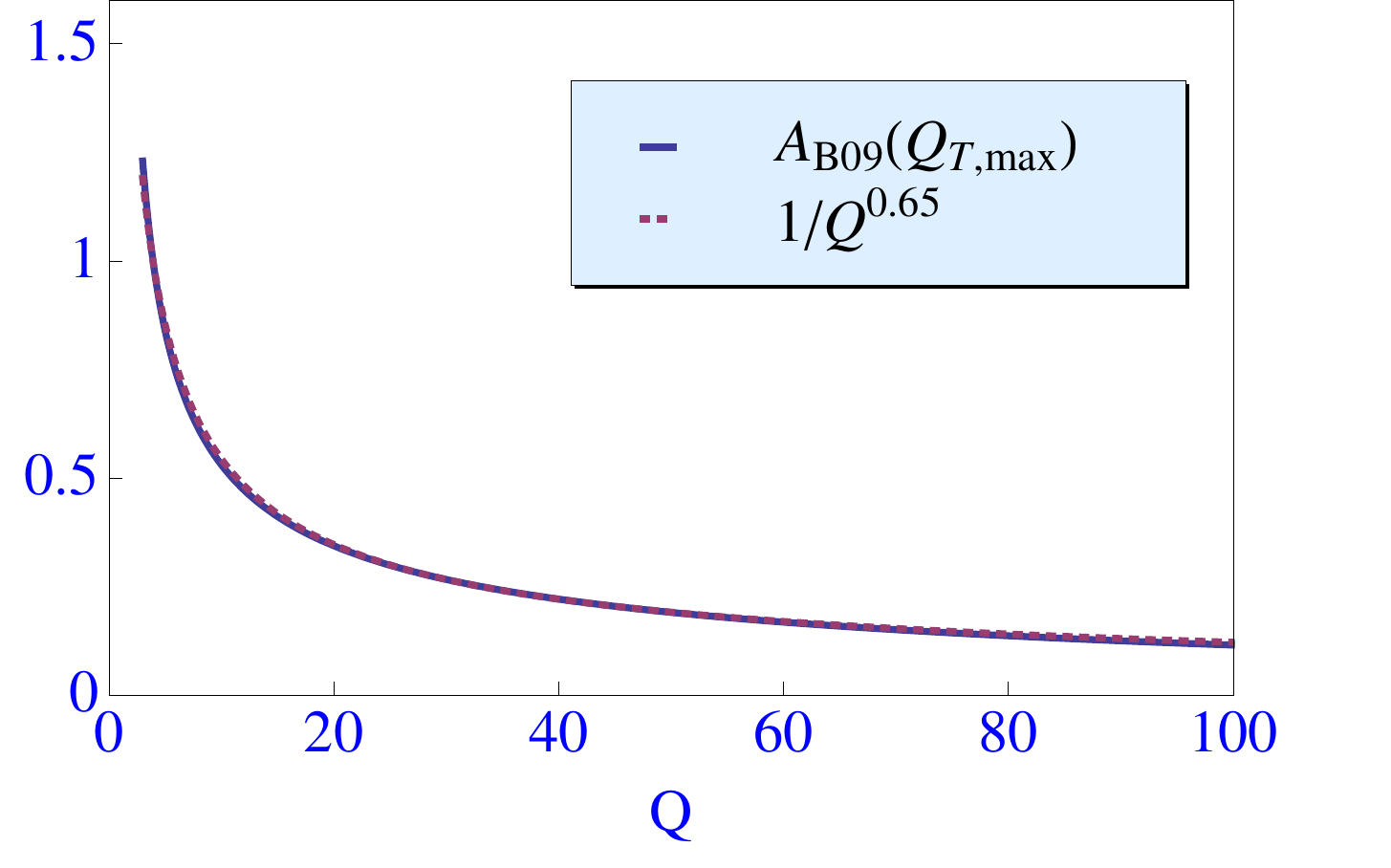}
\caption{Left: the Collins-Soper factorization based asymmetry factor ${\cal A}_{B09}(Q_T)$ (in units of
$M^2$) at $Q=3.33, 10, 30, 60, 90 \, \text{GeV}$. Right: the asymmetry factor ${\cal A}_{B09}(Q_T)$ evaluated at the $Q_{T,\max}$ plotted as function of $Q$ (solid line) and compared to a line (dotted) with a $1/Q^{0.58}$ fall-off, constructed to coincide at the point $Q=30 \, \text{GeV}$.}
\label{SB09plotQT}
\end{center}
\vspace{-2 mm}
\end{figure}

In Fig.\ \ref{SB09plotQT} (left) ${\cal A}_{B09}(Q_T)$ is plotted for various energies. In Fig.\ \ref{SB09plotQT} (right) ${\cal A}_{B09}(Q_{T,\max})$ is plotted as a function of $Q$ and compared to a simple power law approximation $1/Q^{0.65}$ (without soft factor the calculation yields a slower fall-off: $1/Q^{0.58}$). 
Note that the asymmetry is generally smaller than ${\cal A}(Q_T)$ and that the peak of the asymmetry moves more slowly towards higher $Q_T$ values. Nevertheless, the power of the fall-off is very comparable to the result of the previous section. We therefore expect the estimates of the evolution of the $\cos 2 \phi$ Collins asymmetry discussed in B09 (approximately $1/Q$) to not change much when using the recent TMD factorization approach. 

We note that like previously considered perturbative Sudakov factors, $S_{B09}$ also does not have the correct $b\to 0$ limit. 
Due to its $\ln Q_0^2 b^2$ dependence, compared to the $\ln \ln Q_0^2 b^2 $ dependence of $S_p$ in Eq.\ (\ref{SpbQQ0used}), 
the region of very small $b$ (below $1/Q$) contributes more than before. But it turns out to only matter in the denominator of the 
asymmetry to a modest extent (around $Q_{T,\max}$ this region contributes 10-20\% of the denominator of ${\cal A}_{B09}(Q_T)$, 
compared to around 5\% in the case of ${\cal A}(Q_T)$). It may be the reason for a somewhat smaller asymmetry.  
The perturbative tails of the TMDs may matter more in this case. 
These tails contain logarithms of $b_*Q_0$, which only become large in the very small $b$ region. Only if the remainder of the integrand 
has little or no support there, these logarithms 
are of no importance. Given the modest contribution of the very small $b$ region even in this B09 approach, we will not investigate this issue 
further here, but it is an argument in favor of using the new TMD factorization approach as in ${\cal A}(Q_T)$.     
  
\section{Conclusions}
In this paper the energy scale dependence of the Sivers asymmetry in SIDIS has been investigated within the framework of 
TMD factorization. The perturbatively calculable part of the Sudakov factor is considered at the one-loop level, including higher order 
effects due to the running of the coupling constant, in order to avoid the appearance of large logarithms. This study is very similar to the
recent one in Ref.\ \cite{APR} (APR), which focussed on the low energy region $Q<10$ GeV. Here we study a larger $Q$ range and specifically  
the $Q$ behavior of the peak of the Sivers asymmetry, which allows for a direct comparison with earlier results based on Collins-Soper 
factorization in various approximations. Although these treatments differ only beyond the double leading logarithmic approximation, the numerical 
results show that subleading logarithms do matter in the studied $Q$ range. The recent TMD factorization based approach indicates that 
the peak of the Sivers asymmetry falls off with $Q$ approximately as $1/Q^{0.7}$, somewhat faster than was found 
in \cite{B01} ($1/Q^{0.5-0.6}$), but similar to what follows from the approach discussed in \cite{B09} which still includes a soft factor ($1/Q^{0.65}$). 
Since these numerical results involve some approximations, such as neglecting 
the $b$ dependence of the TMDs in the small $b$ region and using the same $S_{NP}$ in both numerator and denominator of the asymmetry,
the actual fall-off to be determined by experiment could be somewhat faster even. From varying the nonperturbative Sudakov factor (also separately 
in the numerator of the asymmetry) we expect a power somewhere in the range $0.6$-$0.8$. Similar moderate differences between the approaches based on the first TMD factorization of Collins and Soper (1981) and on the recent TMD factorization by Collins (2011) are expected also for other azimuthal asymmetries, such as the Collins effect asymmetries studied in \cite{B01,B09}. Of course, this conclusion applies specifically to the studied kinematic range.  

In the numerical results the peak of the asymmetry moves towards higher transverse momentum values as the energy increases, 
by approximately 70\% over the studied $Q$ range of $3 - 100$ GeV. The peak is located at a transverse 
momentum value that is comparable to $Q_0$, where one expects the dominant contribution to the $b$ integral to come from the 
region $b \sim 1/Q_0$, which is the boundary of the perturbative region. Inclusion of the perturbative tails of the TMDs may thus affect 
the location of the peak of the asymmetry. Apart from the power of the fall-off in $Q$, the behavior of the peak as 
a function of $Q_T$ will therefore allow to test the underlying assumptions considered in \cite{APR} and in this paper. 
Clearly, higher $Q$ data on the Sivers asymmetry in SIDIS is required to test the evolution resulting from the TMD formalism, which awaits an EIC.\\ 

\noindent
{\it Note added:} upon completion of this paper, a paper on the energy evolution of the Sivers asymmetries appeared that addresses related topics 
\cite{Sun2013}, in particular questioning the $S_{NP}$ used in APR. As we pointed out, our results for the power of the fall off with energy are not very sensitive to the particular $S_{NP}$ used. In addition, a new study of the effect of TMD evolution of the Sivers function in semi-inclusive $J/\psi$ production appeared recently \cite{Godbole:2013bca}.

\begin{acknowledgments}
The author wishes to thank John Collins and Ted Rogers for very detailed and important comments. 
Furthermore, I thank them, Wilco den Dunnen, Markus Diehl, George Sterman, and Werner Vogelsang for useful discussions, even if they sometimes took place years ago.  
\end{acknowledgments}


\end{document}